# Selective transport of water molecules through interlayer spaces in graphite

Lalita Saini[1], Siva Sankar Nemala[1], Aparna Rathi[1], Suvigya Kaushik[1], Gopinadhan Kalon[1,2]

[1]Discipline of Physics, Indian Institute of Technology Gandhinagar, Gujarat 382355, India

[2]Discipline of Materials Engineering, Indian Institute of Technology Gandhinagar, Gujarat 382355, India

**Interlayer space in graphite is impermeable to ions and molecules, including protons. Its controlled expansion would find several applications in desalination, gas purification, high-density batteries, etc. In the past, metal intercalation has been used to modify graphitic interlayer spaces; however, resultant intercalation compounds are unstable in water. Here, we successfully expanded graphite interlayer spaces by intercalating aqueous KCl ions electrochemically. The water conductivity shows several orders of enhancement when compared to unintercalated graphite. Water evaporation experiments further confirm the high permeation rate. There is weak ion permeation through interlayer spaces, up to the highest chloride concentration of 1 M, an indication of sterically limited transport. In these very few transported ions, we observe hydration energy-dependent selectivity between salt ions. These strongly suggest a soft ball model of steric exclusion, which is rarely reported. Our spectroscopy studies provide clear evidence for cation-π interactions, though weak anion-π interactions were also detectable. These findings improve our understanding of molecular and ionic transport at the atomic scale.**

Introduction

Selective transport of molecules and ions is ubiquitous in nature[1,2]. For example, aquaporins transport water molecules selectively while blocking all the ions, including protons[3]. Mimicking biological channels would result in highly efficient filtration systems. Several low-dimensional materials, including nanopores, nanotubes, channels, and laminates[1,4–8] have been investigated for this purpose. Of these, two-dimensional materials were found to be very attractive due to their thinness, ability to form heterostructures, and the possibility to tune their interlayer spaces[9–12]. However, several challenges limit their potential usage, these include uncontrollable functionalization[13], swelling in aqueous solutions[14], and multiple processes involved in the device fabrication. Moreover, the membrane architecture formed from the assembly of exfoliated individual layers is highly susceptible to large-scale defects and pinholes. To address this problem, we chose high-quality single-crystalline graphite samples. It is reported that, at thermal energies, no ions can permeate through graphite, including small-sized protons[15]. The weak van der Waals interaction of the layers in graphite offers a possibility to modify its interlayer space. The controlled expansion of the layers would allow selective transport of molecules and ions. The intercalation of metal atoms such as lithium, calcium, etc. in graphite results in intercalation compounds[16], which have been well studied. However, these compounds are exothermic in nature and decompose while in contact with water. This prevents their utility in water-related applications.

In our approach, we utilized a geometry and intercalant very different from the conventional intercalation. A thin crystal of graphite is placed in between two reservoirs of aqueous KCl solutions in such a way that the graphite has space to expand upon ion intercalation (Fig. 1a). There are several reasons to choose KCl ions for the intercalation; the hydration radius of $K^+$ and $Cl^-$ is the smallest among all the salt ions[17], and the intercalation is expected to provide an interlayer distance sufficient to let water molecules pass through while rejecting salt ions. The ionic nature of the solution helps the intercalation to be done electrochemically, which is fast, efficient, and a room-temperature process. It is worth mentioning that several salt ions were utilized recently to control the interlayer distance of graphene oxide membranes[12]. However, the intercalation of ions within the graphene oxide membrane is very complex to understand due to the existence of both pristine and oxidized regions, let alone the issue of uncontrolled swelling. The study also investigated the nature of interactions between intercalants and graphene oxide channels[12], especially the cation-π interactions, albeit without much success. Recently, there has been a surge of interest to understand the electric field-controlled ion-transport properties of graphene oxide[18,19]. Our study utilizes electric field to intercalate KCl ions and understand the interaction of KCl ions with graphite surfaces.

**Results**

For the present study, the highly oriented graphitic piece was epoxy glued onto an acrylic sheet that had a pre-fabricated hole of size 2 mm x 2 mm (inset of Fig. 1b). The epoxy ensures that the only path for the ion diffusion is the interlayer space. Before the intercalation process, we tried to measure the leakage current through this sample. Aqueous NaCl solution of concentrations from $10^{-5}$ M to 1 M was used for the measurement. The leakage current for all the concentrations was less than $10^{-10}$ A at an applied voltage of ±100 mV (conductance ~$10^{-9}$ S). Further details about the measurement can be found in the Methods section and the supplementary material (see supplementary section 3).

The as-prepared samples were then taken for electrochemical intercalation in a setup shown in Fig. 1a & supplementary fig. 1. For the intercalation, aqueous solution of 2 M KCl was used. A voltage was applied across the platinum electrodes, which is expected to enhance the ion diffusion through the graphite sample. The diffusion process depends on both the magnitude of voltage and its time duration, and accordingly, we have done several iterations to optimize these values. The applied voltage was varied in the range of 2 - 10 V and the intercalation time from 2 to 12 hours. We *in-situ* monitored the resulting ionic current through the sample, and the successful intercalation is indicated by a sudden increase in ionic current. We also noticed that at this point, graphite begins to exfoliate, and simultaneously the surface becomes rough. A voltage of 10 V for a duration of 3 hours was found to be the optimum for samples of 0.25-1 mm thickness. The ion intercalation is also evident from the samples' recorded X-ray diffraction pattern that shows a new feature at 2θ = 9.10°, corresponding to an interlayer distance of 0.97 nm. In addition, the intense peak (0 0 2) has split into two peaks, and most of the higher angle peaks disappeared, suggesting modification in the graphite crystalline structure (see supplementary fig. 2). To further confirm this modification, we performed water contact angle measurements. Upon intercalation, the graphitic surface becomes hydrophilic as evident from the reduced contact angle of ~70° from ~104° (see supplementary fig. 3). The smaller contact angle after the intercalation process could be partially related to the increased surface roughness. To find the contact angle of inner layers, we removed several layers from the top side of the sample. In this case, the contact angle is ~ 84°, a value close to the reported result from graphene fluidic channels[20]. This measurement suggests that the intercalated graphite (ICG) channel is hydrophilic at the entrance/exist and nearly hydrophobic at the bulk of the channel.

Immediately after the intercalation, the current-voltage (*I-V*) characteristics were measured using equal molar concentration, *C* of NaCl in both the reservoirs. The intercalation clearly shows increased ionic current through the graphitic sample (Fig. 1b). At ±100 mV and a concentration of $10^{-5}$ M, the measured current is of the order of ~$10^{-7}$ A. This concentration roughly corresponds to that of ions in water as the measured pH of our DI water is ~5.5. The corresponding water conductance is ~$10^{-6}$ S, which is three orders of magnitude larger than those measured for unintercalated graphite. The estimated conductance remains nearly constant in the concentration range of $10^{-5}$ - $10^{-2}$ M (Fig. 1c). These measurements were repeated using other salt ions of different cationic hydration diameters and valences. The data for HCl and other salts is shown in figures 1c and 4a, respectively. Different salt ions show similar constant conductance in the concentration range of $10^{-5}$ -$10^{-2}$ M with weak dependency on valence and hydration diameter. Beyond $10^{-2}$ M, the NaCl conductance increases gradually, and at 1 M concentration, the increase is a factor of 10-20 times that of water. Interestingly, at this concentration of 1 M, the conductance of all these salts shows bulk-like behavior as inferred from the ratio of conductances (Fig. 4a). This conductance enhancement through intercalated graphite (ICG) sample is really small when compared to bulk samples where the latter exhibit 5 orders of increase in ionic conductance when the concentration is varied from $10^{-5}$ to 1 M. This strongly indicates high salt rejection efficiency of ICG samples, even at the practically relevant sea salt concentration of 0.6 M. A histogram showing the conductance distribution across several samples is shown in supplementary fig. 4. If the thickness of sample is reduced, an increase in conductance is observed which can be seen in supplementary fig. 4b.

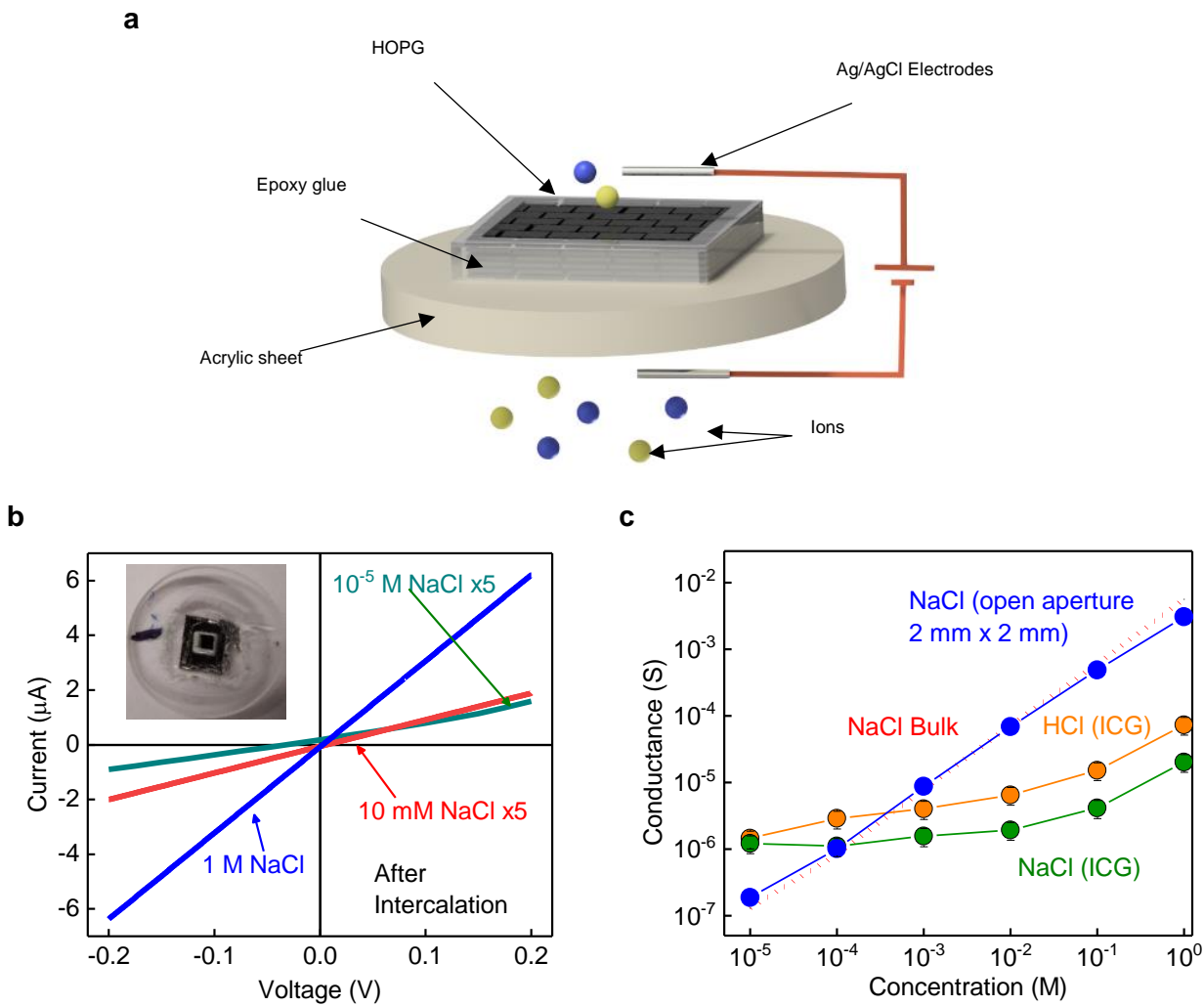

**Fig. 1: Ion-transport through intercalated graphite (ICG). a** Schematic of our ion-transport measurement setup. **b** *I-V* characteristics of ICG with an equal concentration of NaCl in both reservoirs. The top inset shows an actual ICG device. NaCl concentrations vary from $10^{-5}$ M to 1 M. **c** Conductance of ICG for both HCl and NaCl, along with the data from the open aperture support (blue spheres) of area 2 mm x 2 mm and the effective thickness of 0.25 mm. Error bars provide the standard deviation between repeated measurements. The red dotted curve corresponds to standard data taken from literature[31]. The conductance of ICG at 1 M NaCl concentration is ~3 orders of magnitude smaller than that of bulk.

Next, we tried to find out what fraction of ions are transported through the interlayer spaces. We measured the ionic conductance of NaCl through the same acrylic support with a 2 mm x 2 mm hole in it, but this time without the graphite. A comparison of the conductance data with and without graphite shows that at 1 M concentration, the ionic conductance through graphite has been suppressed by 3 orders of magnitude. This suggests that even at the highest concentration of 1 M, the actual ion concentration transported through the sample would be smaller at least by $10^{-3}$ M. We tried to measure the actual concentration of salts that are transported through graphite with the help of diffusion experiments. In this experiment, the two reservoirs, high concentration ($C_H$) and low concentration ($C_L$) were filled with 1 M NaCl and water, respectively. We continuously monitored the diffusion current and a steady state is attained in 24 hours. The diffused ions from the $C_L$ side, if any, were collected (inset of Fig. 2b) and examined using inductively coupled plasma (ICP)-mass spectrometer (MS). The osmotic pressure forces an equivalent amount of water from the $C_L$ side to $C_H$ side. Therefore, this measurement provides an upper-bound estimate on the Na concentration, which is ~18 ppm. This is equivalent to 7 x $10^{-4}$ M of Na, which agrees with the result of the conductance measurement. The volume change in our case is rather small for a duration of 24 hours[21], which allowed us to estimate the salt rejection efficiency as 1- $C_L$ / $C_H$, which is more than 99%.

To find out more about the role of interlayer spaces in the transport, we performed drift-diffusion measurements. The reservoirs were filled with different concentrations of aqueous KCl solutions. We choose KCl due to similar diffusivities of $K^+$ and $Cl^-$ that ensures zero contribution from liquid junction potential to the measured potential. In these measurements, the concentration of $C_H$ was fixed to 1 M, and the $C_L$ to 1 mM, 10 mM, and 100 mM and 1 M. In the absence of any applied voltage, any difference in the diffusion rates of cations and anions appears as a finite current, consequently, the *I-V* curves shift along the voltage axis (Fig. 2a). For our configuration (Inset of Fig. 2b), the negative current at zero potential suggests higher diffusion rates for cations compared to anions. The potential corresponding to zero current, $V_0$ is estimated from the *I-V* curves (Fig. 2a). We estimated the diffusion potential, $V_{diff} = V_0 - V_R$, where $V_R$ is the redox potential. $V_{diff}$ is found to exhibit a logarithmic relationship with the concentration ratio (Fig. 2b), which can be described by the Nernst equation[22]. This allowed us to estimate the ion selectivity, *S* as described below

$$V_{diff} = S\frac{RT}{F}\ln\left(\frac{C_H}{C_L}\right) \quad (1)$$

Here, $S = t_+ - t_-$, where $t_+$ and $t_-$ are the transport number of cations and anions, respectively. $S = 1$, for the ideal cation-selective channels and $S = 0$, for the non-selective channels. $R$, $T$ and $F$ are the Universal gas constant, the temperature and the Faraday constant, respectively. The selectivity in our case is found to be 0.26, which suggests weak selectivity between cation and anions. This observation along with the constant conductance at lower salt concentrations, at first instance, might probably indicate a constant surface charge governed regime, however as we discuss later, the constant conductance is most likely arising from steric exclusion.

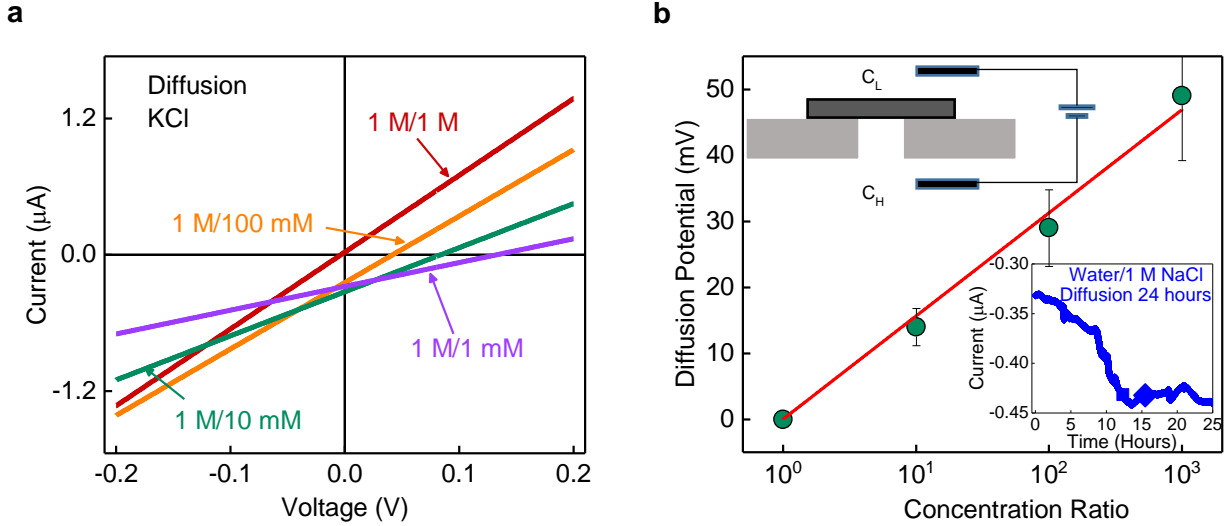

**Fig. 2: Poor ion selectivity of intercalated graphite. a** Current-voltage characteristics for various concentrations of KCl under a concentration gradient. The concentration of one of the reservoirs ($C_H$) was fixed to 1 M and the other ($C_L$) was varied from 1 M to 1 mM. **b** The estimated diffusion potential is plotted as a function of the concentration ratio. The red solid line indicates the best fit to the measured data. The error bar provides an estimate on the spread in the diffusion potential in repeated measurements. The top inset shows the schematic of the drift-diffusion experiments. The bottom inset shows the diffusion current measured over a duration of 24 hours, with water on one side of the sample and 1 M NaCl on other side.

To understand the nature of the transport of water molecules through interlayer spaces, we performed evaporation experiments using a high-precision gravimetric setup (see supplementary fig. 5). The schematic of the setup is shown in Fig. 3a. The weight loss as a result of evaporation through the sample is monitored as a function of time. No significant weight loss was detected through unintercalated graphite samples, for more than 12 hours (Fig. 3b). On the other hand, the intercalated graphitic samples showed large weight loss sufficiently larger than those measured from a similar sized open aperture. To quantify the evaporation rate, $Q$, we determined the slope of the weight loss with respect to time. For the intercalated graphite samples that we used in our ion-transport measurements, $Q \approx 1$ µg s$^{-1}$. We observed that $Q$ is highest and more or less constant if the samples are in direct-contact with water. We found no effect on $Q$ for different levels of relative humidity. The active area, $A$, that is responsible for $Q$ was estimated from the ionic conductivity data, which comes out to be 3.4 mm$^2$, close to that measured using an optical microscope. Unlike previous studies, our water conductivity experiments allowed us to estimate the effective area very accurately. We utilized this $Q$ to find the flux by considering the active area of the samples, which is ~1 L m$^{-2}$ h$^{-1}$. We additionally performed forward osmosis experiment, using sucrose on one side of the ICG and water on the other side to estimate the water permeation flux. The measured volume change and the estimated water flux comes out to be very similar to that calculated from water evaporation measurements. This is not surprising given that in both experiments, water flows as a liquid. More details on the forward osmosis experiment is provided in supplementary section 7.

Several studies on water flow through narrow hydrophobic carbon nanostructures indicate that water remains in liquid state inside these structures[20,23,24]. We checked this possibility in our intercalated graphite samples with the help of Hagen-Poiseuille equation with slip-boundary conditions. We estimated the pressure difference ($\Delta P$) that is responsible for the high evaporation rates as

$$\Delta P = \frac{12Q\eta L}{\rho h^3 w}\left[1 + \frac{6\delta}{h}\right]^{-1} \qquad (2)$$

Where $\eta$ and $\rho$ are the viscosity and the density of water, respectively. $L$, $w$ and $h$ are the length, width and the interlayer spacing, respectively. Here $L$ = 0.25 mm, w = 2 mm. The channel height ($h$) that is available for the passage of water molecules is taken to be 6.3 Å, which is obtained by subtracting the thickness of the carbon sheet (3.4 Å) from the measured interlayer spacing (9.7 Å). $\delta$ is the slip length, which is taken to be 10 nm, as reported for carbon nanotubes[20,23]. The areal density of the channels is estimated from the relation $A/(w \times h)$. The estimated pressure is ~65 bar, a value smaller than previous reports[20,24]. High water flux is also reported in rGO membranes with an interlayer spacing of 0.37 nm[25], where the fast flow is attributed to hydrophobic graphitic regions. We also observe a very similar fast flow through our samples, albeit with high salt rejection. We, however, stress here that high water evaporation rate was not observed in the unintercalated or pristine graphite samples, which clearly indicate the importance of intercalation in our intercalated samples.

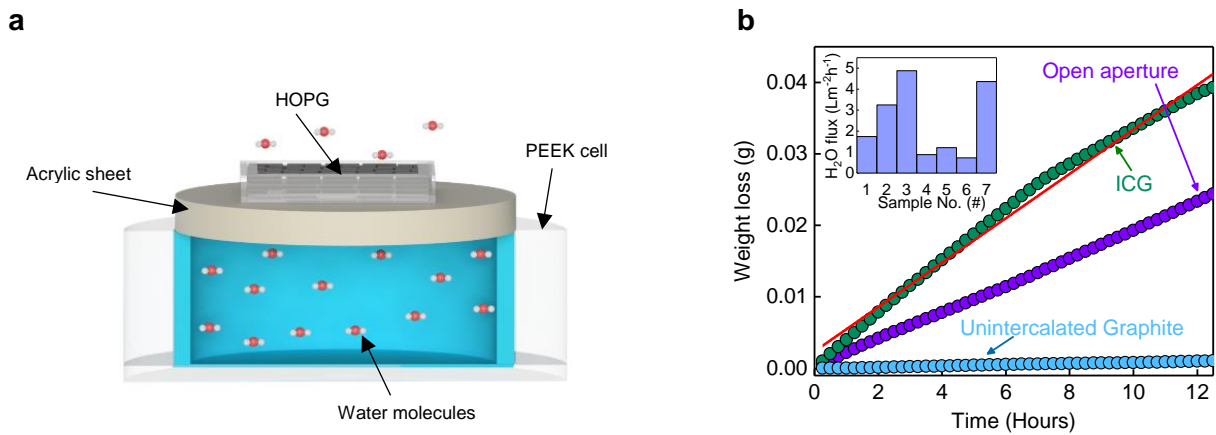

**Fig. 3: Water permeation through intercalated graphite. a** Schematic of our water evaporation setup. **b** Weight-loss through graphite samples is monitored as a function of time, which is a measure of the water evaporation rate. The area of the sample is 3.4 mm$^2$. Over a span of 12 hours, the measured weight-loss was little for the unintercalated graphite sample, whereas the intercalated graphite sample shows weight loss more than an open aperture. The solid red line is the best fit to the measured data. Inset: water evaporation flux measured from seven such samples of ICG.

To confirm that potassium and chlorine are indeed present in the interlayer spacing, EDAX-SEM analysis of the samples were done, both before and after the intercalation. The pristine graphite sample was of high purity and as expected we could detect only carbon before the intercalation process. Immediately after the intercalation, we performed cross-section analysis of the graphite sample. For this, the sample was washed several times with deionized water and IPA for removing any residual salt that is on the surface. After this, the sample was sliced from the middle and the middle surface was utilized for the cross-section analysis. In the EDAX-SEM, we could detect both potassium and chlorine in the intercalated graphite sample, in addition to the parent carbon (see supplementary fig. 6). An elemental mapping indicates that both K and Cl are uniformly distributed over the surface.

To find out if there is any chemical change in the graphite structure due to intercalation, X-ray photoelectron spectroscopy (XPS) analysis of samples was carried out. In the graphite samples that were not intercalated, we mostly detected carbon (91.3 atomic %) along with some oxygen (8.7 atomic %) (see supplementary fig. 7), though the origin of oxygen is not very clear. After the intercalation of KCl, carbon (91.89 atomic %), oxygen (2.64 atomic %), potassium (4.77 atomic %) and chlorine (0.7 atomic %) could be detected in the graphite sample. XPS was also carried out at different depths (see supplementary fig. 8) to understand the distribution of potassium and chlorine across the sample. We found that the amount of salt ions is very similar at different depths, indicating its uniform distribution, in agreement with EDAX results. This result further strengthens our observation that graphite intercalation is successful. The presence of potassium is more favorable inside the graphite due to cation-π interactions[12,26], however we could still detect chlorine which might be an evidence of less-discussed anion-π interactions[26].

A very recent DFT calculation[27], predicts that both ions are favorably incorporated in the interlayer spaces, which agrees with our experimental observation. The samples are highly water stable without any evidence of swelling and the characteristics were unaffected even after storing in water for several months. The results are highly reproducible, though slight variations in conductance between different samples were observed due to variations in the geometrical parameters.

Having discussed the stability of these samples, we will now discuss in detail the transport characteristics at higher salt concentrations. At a concentration of 1 M, we observe low but distinct conductance for various salts (Fig. 4a), although the actual concentration of the ions transported is smaller than $10^{-3}$ M. Such a low concentration of the transported ions suggest that the ions are rejected due to steric exclusion. The conductance data on various salts at 1 M concentration, interestingly allow us to probe further the nature of the steric exclusion. The small but finite ionic conductance probably already indicates the soft nature of the hydration shells instead of the commonly discussed hard ball model. We observe that the conductance of the salts decreases in the order; KCl > NaCl > $CaCl_2$ > $MgCl_2$. The conductance between salts shows a much larger difference than expected for bulk solutions. For example, in our samples, the conductance ratio of KCl and $MgCl_2$ is ~ 3.0, whereas in bulk, it is only ~1.4. The data suggests that the conductance of the ions in interlayer spaces is most likely controlled by its (de)hydration energy barrier. To confirm this, we estimated the conductance of salts with respect to HCl conductance and plotted this conductance ratio on a logarithmic scale along with the cation (de)hydration energy of different salts, which is shown in Fig. 4b. Here, the (de)hydration energy is the energy required to completely remove the hydration shells of an ion. We clearly observe a one-to-one correspondence between salt conductance and its (de)hydration energy, and hence further confirm the soft nature of the hydration shells in agreement with previous reports from Jain *et al.*[4] and Esfandiar *et al.*[28]. These observations strongly argue that at lower concentrations, the observed constant ionic conductance is most likely a result of steric exclusion and not related to surface charge.

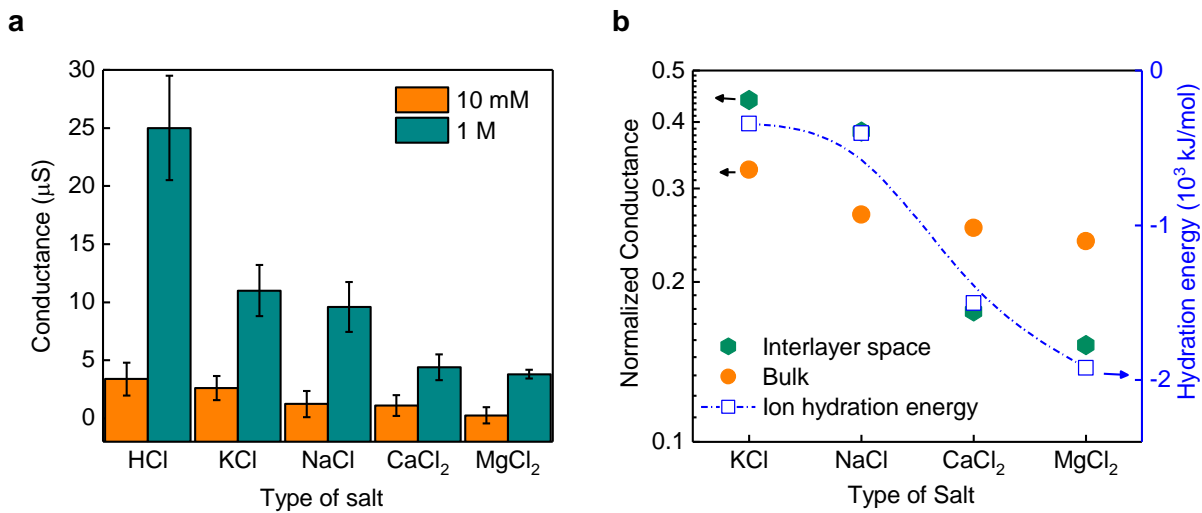

**Fig. 4**: **Steric exclusion and soft type transport through intercalated graphite. a** Conductance of several salts for concentrations of 10 mM and 1 M (error bar and is calculated from the data of 4 different samples). At lower concentrations, all the salts show very similar conductance, however at higher concentrations (1 M), the conductance decreases with increase in (de)hydration energy of cations. **b** Plot of conductance normalized with respect to HCl for various salt solutions, through ICG and an open aperture. The (de)hydration energy of cations, taken from literature[17] is plotted along the right Y-axis as indicated by blue square symbols. Here, (de)hydration energy is the energy required to completely remove the hydration shells of an ion. The blue dotted line is a guide to the eye. There is one to one correspondence between salt conductance and (de)hydration energy.

This result also suggests that it is nearly impossible to achieve 100% salt rejection and the soft nature of the ions impose a fundamental limitation. Experimental demonstration of steric exclusion is rarely reported[4,8,28], as it requires fluidic channels of dimensions smaller than ion hydration sizes, where the latter is typically larger than 6.6 Å. The demonstration of steric exclusion in our case is a result of the smaller interlayer spaces with a height of ~6.3 Å. Our study thus provides a foundation for ion separation that relies on hydration energies.

The estimated water flux of ~1 L m$^{-2}$ h$^{-1}$ is comparable to 5-10 L m$^{-2}$ h$^{-1}$ that typically achievable in the forward osmosis set up. It is worth mentioning here that the thickness of our sample is only 0.25 mm, and there is plenty of room at the bottom to improve the water flux. We also observed that several samples showed water flux in the range of ~1-5 L m$^{-2}$ h$^{-1}$ (inset of Fig. 3b), though we confined our discussion to those samples that showed the lowest and exhibited the most reproducible transport properties among them. With high-water flux rate, stability and greater than 99% salt rejection efficiency makes the intercalated graphite an ideal candidate for desalination applications.

To better understand the role of KCl ions in enhancing the ionic conductance through interlayer spaces of graphite, we used deionized (DI) water as the intercalant instead of aqueous KCl. We kept all the intercalation parameters same as that of KCl. With water as the intercalant, there was only a little increase in ionic current through the graphite samples (see supplementary fig. 9). The small enhancement hints toward the importance of ions in the intercalant solution. The presence of $H_3O^+$ and $OH^-$ ions in DI water seems to help the intercalation process, though with very little efficiency, even at the highest applied voltages of 10 V. Visibly, water intercalated samples had a smooth surface morphology in comparison to KCl intercalated samples, where the latter showed a rough surface after the intercalation process. This experiment also helped us to completely rule out the possibility of any contribution from defects that might be created as a result of high applied voltages. Further, high water flow rates suggest the importance of hydrophobic surfaces, and also the higher water conductance and salt rejection suggests very little possibility of defects dominating the transport.

Having discussed the transport of water molecules and ions through millimeter-sized graphite samples, we now discuss several approaches to its scalability. As the basic building block of these transport channels are graphene layers, thin and large area graphene membranes can be prepared via liquid-phase exfoliation in suitable solvents, followed by vacuum filtration or spin coating. Several studies have already indicated this possibility[29,30]. This approach is very similar to graphene oxide (GO) membrane fabrication. Additionally, graphene membranes can also be prepared by transferring single-layer CVD-grown graphene on top of each other, leading to the possibility of wafer-scale devices.

In summary, an alternate and successful route for the electro-chemical intercalation of KCl in graphite is demonstrated that yields high salt rejection efficiencies and enhanced water permeation rates. XPS and EDAX-SEM analysis clearly showed homogeneous distribution of K and Cl inside graphitic interlayer spaces with evidence for dominant cation-π interactions along with non-negligible anion-π interactions. The weak cation vs anion selectivity and low salt conductance is a clear indication that the transport is sterically limited. Bulk-like transport at 1 M concentration suggests a soft ball model of ion transport and the selectivity among salts is determined by their hydration energy. The observed water evaporation rate is successfully explained based on the Hagen-Poiseuille equation for liquid water flow, with slip boundary conditions. High water to ion selectivity and mechanical stability make ICG very useful as a barrier film for filtration processes, and dehumidification applications.

**Methods**
**Ion transport measurement**
The small graphitic piece was glued (Stycast 1266 Epoxy, part A and B) onto an acrylic sheet that had a pre-fabricated hole of size 2 mm x 2 mm. A Keithley 2614B source meter and Ag/AgCl electrodes were used for the purpose of current measurements. Reference electrodes (HANA instruments, USA) were also used to verify the measured results.
**Water evaporation measurement**
The measurement setup consists of intercalated graphite samples epoxy sealed on top of a miniature container filled with water. The whole setup was placed in a METTLER TOLEDO high precision balance and the weight loss was monitored as a function of time. The precision of the setup is 10 µg. The relative humidity (*RH*) inside the balance was controlled to < 20% in all the measurements.

## Supplementary Information

**Section 1. Sample Preparation:**

For the intercalation, we have used a block of highly oriented pyrolytic graphite (HOPG) (size 12 mm x 12 mm x 2 mm) commercially obtained from Materials Quartz, Inc, USA. The as-obtained HOPG was then carefully cut into smaller pieces with sizes 6 mm x 6 mm x 0.25 mm with the help of a scalpel blade. This was then transferred onto an acrylic sheet that has a pre-prepared square hole of size 2 mm x 2 mm. This acrylic sheet serves as the mechanical support for the transferred HOPG, and the HOPG is tightly sealed with the help of Stycast epoxy glue (inset of Fig. 1b). The samples were left to dry for a minimum of 12 hours, which is the typical cure time for the epoxy setting. After this, we accurately measured the actual transport area of the graphite with the help of an optical microscope (Nikon Eclipse E200). We also tried several other polycrystalline graphite samples; however, those samples were found to be very leaky and exhibit very high ion currents with very small applied voltages, making them unsuitable for the intercalation study.

**Section 2. Wetting and cleaning procedures:**

We wetted the surface of the sample at the beginning of each measurement. For this, we immersed the sample in 2-Propanol (IPA) for about 30 minutes. Prior to each set of measurements, the electrochemical cell was washed sequentially with IPA (100), IPA + DI water (50:50) and DI water (100) for proper wetting of the sample and removal of any residual salts. Ion transport measurements that utilize different salt concentrations the measurements were always taken from low concentrations to high. Between the measurements with different salt solutions, the cell was rinsed with DI water multiple times until the current observed through graphite was equal to the initially measured current for DI water. This data also helped us to monitor the stability of the devices. We found that in the successfully intercalated samples, more than 90% of the devices were water stable with no indication of any degradation in the transport characteristics. These sub-nm channels are very sensitive to molecules' adsorption, which leads to blockage of channels in the dry state, so it is essential to keep the samples in water all the time.

**Section 3. Ion-transport studies:**

An electrochemical cell was custom-made with PEEK (polyether ether ketone) material for the ion transport studies. The cell has two reservoirs with a capacity of 5 ml each. The sample is placed in the middle of two reservoirs, held firmly on two O-rings on both sides of the sample (Supplementary fig. 1). The cell was thoroughly cleaned with acetone, DI water, and IPA in an ultra-sonicator bath for 30 minutes each, before mounting graphite for intercalation. This cleaning procedure is very important to avoid the formation of any air bubbles inside the cell. After thoroughly wetting the sample surface, salt solutions were filled in two reservoirs such that the graphite surface was fully immersed in the solution. The acrylic support and epoxy glue ensured that graphite is the only path for the ions or water molecules. We measured the leakage current of our setup using a blank acrylic sheet. The detected leakage current is ~$10^{-12}$ A for a maximum applied voltage of 200 mV. This current is several orders of magnitude smaller than the recorded values of current through our intercalated graphite samples.

To further check the possibility of any leakage at the interface of an acrylic base and epoxy glue, a voltage similar to that used for intercalation (10 V) was applied to an acrylic piece pasted on a 2 mm x 2 mm hole. The measured current through this sample was still ~$10^{-12}$ A, which is also the lower limit of our measurement setup. This means that the epoxy interface is stable even after applying a high voltage without any generation of defects.

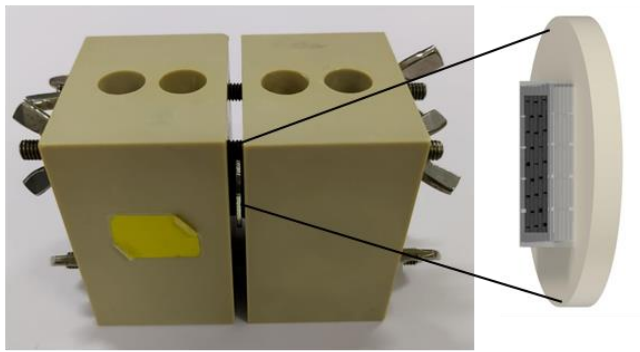

**Supplementary fig. 1: Electrochemical cell.** The optical image of the custom-made electrochemical cell along with an enlarged view of the graphite sample on an acrylic support.

To estimate the salt conductance, we performed *I-V* measurements and the slope of the curve is taken as the measure of conductance. For this, a voltage is swept in the range of 200 mV, in steps of 25 mV and the resultant current is recorded. At low salt concentrations, we observed a small hysteresis in the *I-V* loop measurements. We could remove this hysteresis by increasing the applied voltage duration from the typical 5 s to 180 s, so that the accumulated charge is completely discharged. The small current through uninterialated graphite indicates high quality of our graphite sample and also the efficacy of epoxy sealing. We note that a similar leakage current was reported in the case of ion transport through blocked BN nanotube assembly[1]. Just after this measurement, we proceeded with the intercalation of 2 M KCl. For this, we have filled both reservoirs with 2 M KCl, and applied a maximum voltage of 10 V for a duration of maximum 3 hours for thick samples. For very thin samples (~30 μm), the optimum time was found to be hundreds of seconds for an applied voltage of 10 V. After the completion of intercalation, the assembly was thoroughly cleaned, which removed any residual salts present at the surface.

The success of the intercalation was inferred from the (i) enhanced ionic current of 1000 times to that of unintercalated samples, (ii) A new peak in the XRD pattern at lower angles, (iii) presence of K and Cl in the SEM-EDAX mapping, (iv) presence of K and Cl in the XPS spectra of intercalated samples (v) increased water flux. Graphite intercalation was also done with other reported methods[2] and compared against the XRD patterns obtained for our method (Supplementary fig. 2), including $H_2SO_4$ intercalation of graphite using the secondary intervening method as described in[3]. The sample intercalated with $H_2SO_4$ did not significantly change the XRD peak intensities though the sample was visibly expanded. This suggests that the success of intercalation is tough to be judged with the XRD technique alone. We also used the electrochemical intercalation method where the graphite is made as one of the electrodes with 2 M KCl as the electrolyte and platinum as a counter electrode. We observed that the graphite electrode expanded very quickly (hundreds of seconds) in this experiment.

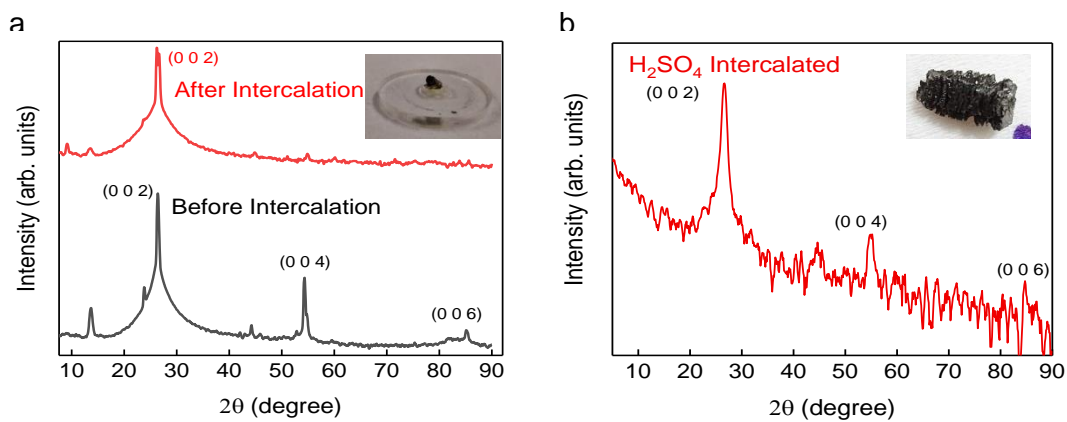

**Supplementary fig. 2**: **The effect of interlayer distance upon intercalation.** (a) X-ray diffraction (XRD) plot of graphite before and after the KCl intercalation. A new peak at 2θ = 9.10° emerges after the intercalation, suggesting successful intercalation of KCl. Inset shows a graphite device that has been intercalated as indicated from the visible expansion. In addition, several intense peaks at higher angles almost vanished after the intercalation process. (b) XRD pattern of $H_2SO_4$ intercalated graphite. The inset shows the visibly expanded sample as a result of intercalation.

We have performed contact angle measurements to clarify the surface modifications and the water intake mechanism (Supplementary fig. 3). The applied voltage helps the entry/exit of the graphite interlayer space become hydrophilic, which helps the water enter the interlayer space via capillary action. The nearly hydrophobic bulk channel induces slip-enhanced water flow.

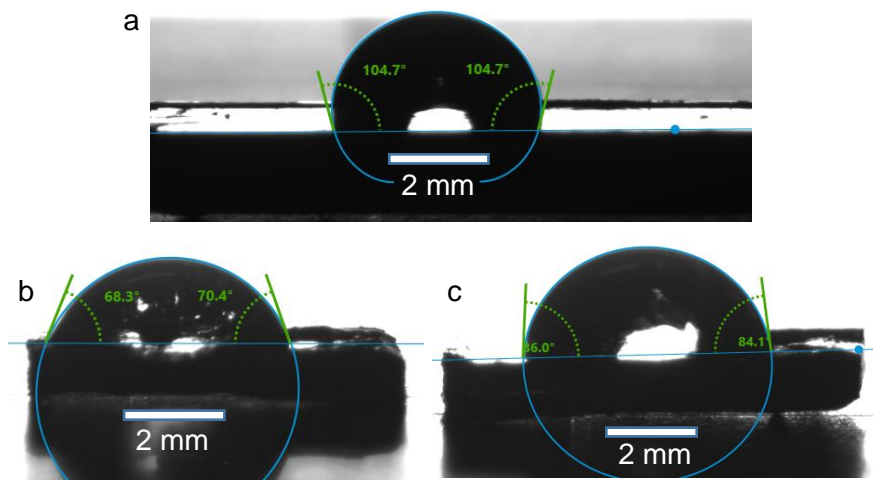

**Supplementary fig. 3: Contact angle measurements.** (a) The graphite surface is hydrophobic before the intercalation. (b) After the intercalation, the graphite surface becomes hydrophilic. (c) After removing a few top layers of intercalated graphite, the surface is less hydrophobic than the pristine sample.

The repeatability of the results was checked by intercalating approximately 14 samples of similar areas. A histogram showing the variation in water conductance of different samples is shown in supplementary fig. 4. The water conductance lies in the range of 0.2-2.4 x $10^{-6}$ S.

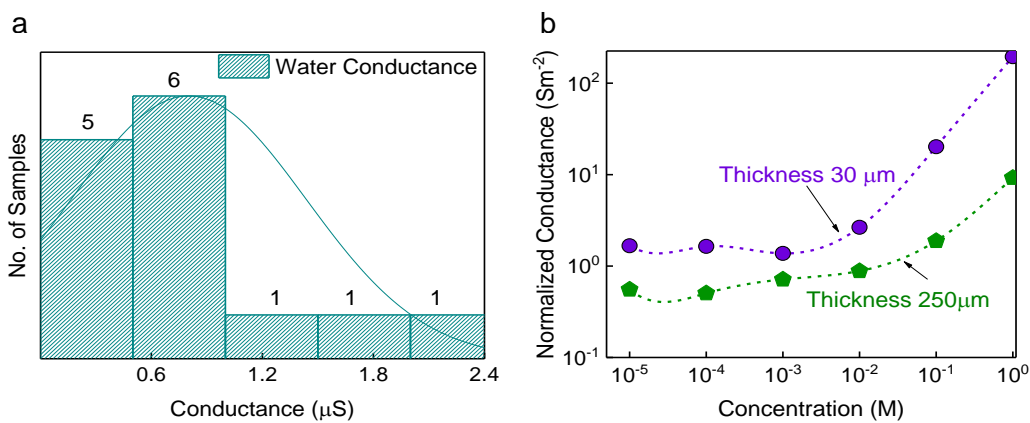

**Supplementary fig. 4: Reproducibility of the samples.** (a) Histogram showing the water conductance of 14 samples of similar geometrical dimensions after the intercalation. Above each rectangular bar, the sample number is indicated. (b) NaCl conductance (normalized for area) variation with thickness.

**Section 4. Water evaporation and weight-loss studies:**

After the ion-transport studies, the same sample is used for measuring the water evaporation rate. For this, we cleaned the sample with 2-Propanol to remove any salt residues on the surface. After that, the sample is placed inside a liquid cell (Supplementary fig. 5). This cell is made up of PEEK (Polyether ether ketone) and the sample is placed in between two O-rings, which keeps the sample leak-tight. The water evaporation assembly is placed inside a high precision micro balance (Mettler Toledo XSR105) of resolution, 10 µg.

A LabVIEW program connected to the precision balance is used to record weight loss for more than 12 hours, with a 1-minute gap between the readings. For controlling the humidity level, silica gel was heated for 10 minutes under an IR lamp and then kept inside the precision balance, well before the measurements. The same procedure was used to measure the weight loss through all the samples, including ICG, 2 mm x 2 mm hole (reference), and unintercalated graphite sample. In majority of the cases, the weight loss rate through ICG was found to be higher than the reference sample. In very few samples, a smaller weight loss than the reference was also observed within 4 to 8 hours when the water was accidently not in contact with the surface of the sample. For unintercalated graphite, the weight loss was significantly lower than the reference.

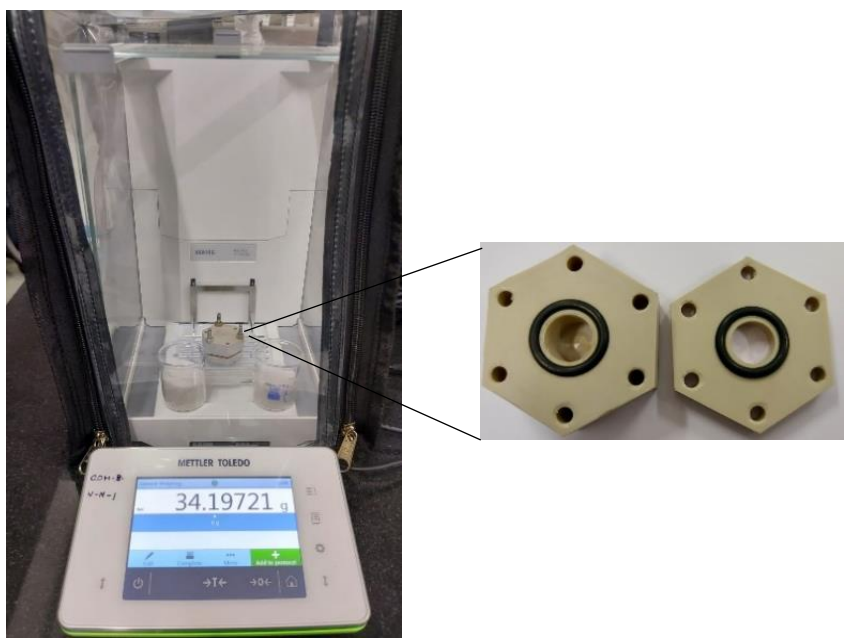

**Supplementary fig. 5: Assembly used for the measurement of water evaporation rate.** The evaporation rate is estimated from the recorded weight loss at different intervals of time, measured using a precission weighing balance (left image). The liquid cell that is used for the measurement is shown in the right image.

**Section 5. Estimation of weight loss through open aperture:**

For ensuring the accuracy of our gravimetric setup, we prepared a reference sample with 2 mm x 2 mm hole drilled through a 2 mm thick acrylic sheet. We measured the water evaporation rate through the open aperture using the gravimetry setup discussed above. For our aperture, the Knudsen number is small and we can safely assume that the water evaporation can be described by diffusion of water molecules through air. The molecular flow is given by[4]

$$F = \frac{1}{3} <v> l \frac{dn}{dx} \quad \quad \text{Supplementary equation (1)}$$

Where $<v>$ is average velocity of water molecules in air, taken to be 590 m s$^{-1}$; $l \approx$ 60 nm is the mean free path and dn/dx is the concentration gradient. When leaving the container, water molecules diffuse through air for a distance equal to the thickness of our reference sample (2 mm). Thus, we can write dn/dx = Δn/t. Where *Δn* is the difference in water concentrations at large distances from the aperture and *t* is the thickness of our reference sample. Further, *Δn* can be estimated as ΔP/k$_B$T; where *ΔP* is difference in partial pressure, *k*$_B$ is Boltzmann constant and *T* is temperature. For solving the diffusion problem exactly, we need to choose between the limit of a steady state of thin film or un-steady state of thick slab[5]. A simple criteria to choose between the two is given in[5], according to which if the ratio

$$\frac{(Length)^2}{(diffusion\ coefficient)(time)}$$

is much less than unity, we can assume that the system is in a steady-state. The measurement time of our evaporation experiment is typically longer than 24 hours, so we can consider this as a steady-state diffusion process, and hence it boils down to the diffusion problem for the case of thin orifices[5]

$$\frac{dn}{dx} = \frac{4}{\pi}\left(\frac{\Delta n}{t}\right) \quad \quad \text{Supplementary equation (2)}$$

The resulting weight loss is

$$Q = FM_{H_2O} A \quad \quad \text{Supplementary equation (3)}$$

where $M_{H_2O}$ is the molar mass of water and *A* is the area of the open aperture. Using *ΔP* as 23 mbar, from the above equation, we get $Q \approx$ 5.02 x 10$^{-7}$ g s$^{-1}$, which is in good agreement with the experimentally measured value through the open aperture, indicating the accuracy of our experimental setup.

**Section 6. Comparison with classical flow equations:**

If we assume that water behaves as a classical liquid inside the interlayer spacing, then we can apply the Hagen-Poiseuille equation to estimate the flow rate as

$$Q \approx \frac{h^3 \rho}{12\eta}\left(\frac{\Delta P}{L}\right) w \qquad \text{Supplementary equation (4)}$$

where $\rho$ and $\eta$ are the density and the viscosity of water, taken as 998 kg m$^{-3}$ and 1 mPa s respectively and $L$ is the thickness of the sample, which is estimated to be 0.25 mm. $w$, is the width, which is taken as ~2 mm. The areal density of the channels is estimated from the relation $A/(w \times h)$. With vapor pressure as 23 mbar, the estimated flux is ~4×10$^{-7}$ L m$^{-2}$ h$^{-1}$, which is seven orders of magnitude lower than what is observed experimentally. But, if we consider the interaction between water and graphite edges, the effective pressure is assumed to be of capillary in origin, and hence can be taken as 1 bar, just to get a rough idea. The water flux estimated is ~14 µL m$^{-2}$ h$^{-1}$, four orders of magnitude smaller than the measured value. So, in agreement with the previous reports, a slip correction term must be included in the Hagen-Poiseuille equation due to the hydrophobic nature of graphite walls, which is discussed in the main text.

**Section 7. Forward osmosis experiment for the estimation of water flux:**

Forward osmosis uses osmotic pressure gradient to draw water molecules across a semi-permeable membrane. In this process, draw solution drives water molecules across the sample from feed solution. Our study used equal amounts of 2 M sucrose and deionized water as draw solution and feed solution, respectively, separated by a sample with an effective area of 3.4 mm$^2$ and thickness of 0.25 mm. The osmotic pressure ($\pi$) is calculated from the van't Hoff equation

$$\pi = \phi i R T M \qquad \text{Supplementary equation (5)}$$

where M, R and T are molar concentration (mol L$^{-1}$), universal gas constant and temperature, respectively. $\phi$ is osmotic coefficient ($\varphi_{sucrose}$ = 1.02), and $i$ is the number of ions in which the solvent dissociates ($i_{sucrose}$ = 1). For our setup, these values provide an osmotic potential gradient of ~50 bar. The actual values of pressure might be slightly different due to capillary pressures. We have observed a 100 µL increase in volume after 24 hours, which translates into a water permeation rate of 1.22 L m$^{-2}$ h$^{-1}$, a value close to that estimated from water evaporation experiments. We have utilized high-precision micro-pipettes to measure small volumes with reasonable accuracy. We also observed changes in water levels from the captured images and method followed in[6]; however, was difficult to estimate small changes in volume accurately.

**Section 8. Confirmation of KCl intercalation with SEM-EDAX and XPS:**

EDAX-SEM analysis of the samples was also carried out, both before and after the intercalation process (Supplementary fig. 6). For this, the samples were sliced from the middle and the cross-section was analyzed. Only carbon was detected in the sample before the intercalation, while after the intercalation, potassium and chlorine were also detected. A complete mapping of K and Cl shows that it is uniformly distributed across the layers.

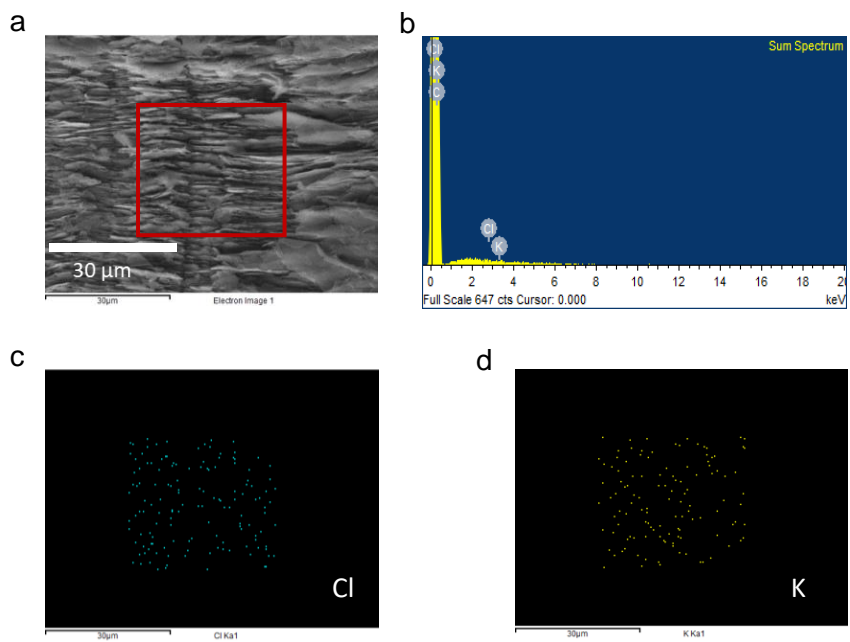

**Supplementary fig. 6: SEM-EDAX characterization of samples after intercalation.** (a) The cross-section of intercalated graphite. (b) EDAX peaks show the elements present in the area enclosed by the red box in (a). Elemental mapping indicates the presence of chlorine (c) and potassium (d) in the sample.

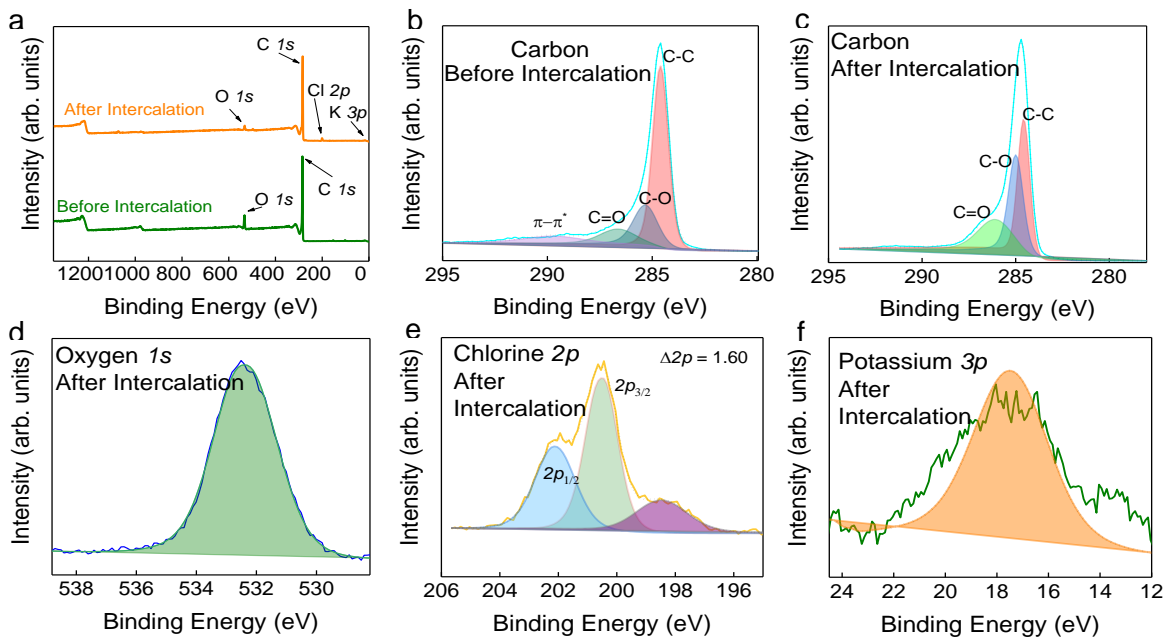

**Supplementary fig. 7: X-ray photoelectron spectroscopy (XPS) results.** (a) XPS survey of graphite sample before and after the intercalation of KCl. The presence of chlorine and potassium is evident in the intercalated samples, while this was absent in the unintercalated samples. The carbon peaks look similar both before (b) and after (c) the intercalation process, except for the absence of π-π* peak after the intercalation process. The presence of oxygen (d) was detected before and after the intercalation process, though the percentage is smaller after the intercalation. Fully resolved chlorine (e), and potassium (f) peaks in the intercalated sample allowed us to estimate the atomic percentage of potassium and chlorine as 4.77 and 0.7, respectively.

We also performed X-ray photoelectron spectroscopy (XPS) (Make, Thermo scientific and Model, ESCALABE 250Xi and Al source) analysis of the samples before and after the intercalation of graphite (supplementary fig. 7). The presence of K and Cl inside the layers is clearly evident in the intercalated graphite samples and are uniformly distributed across the depth (supplementary fig. 8).

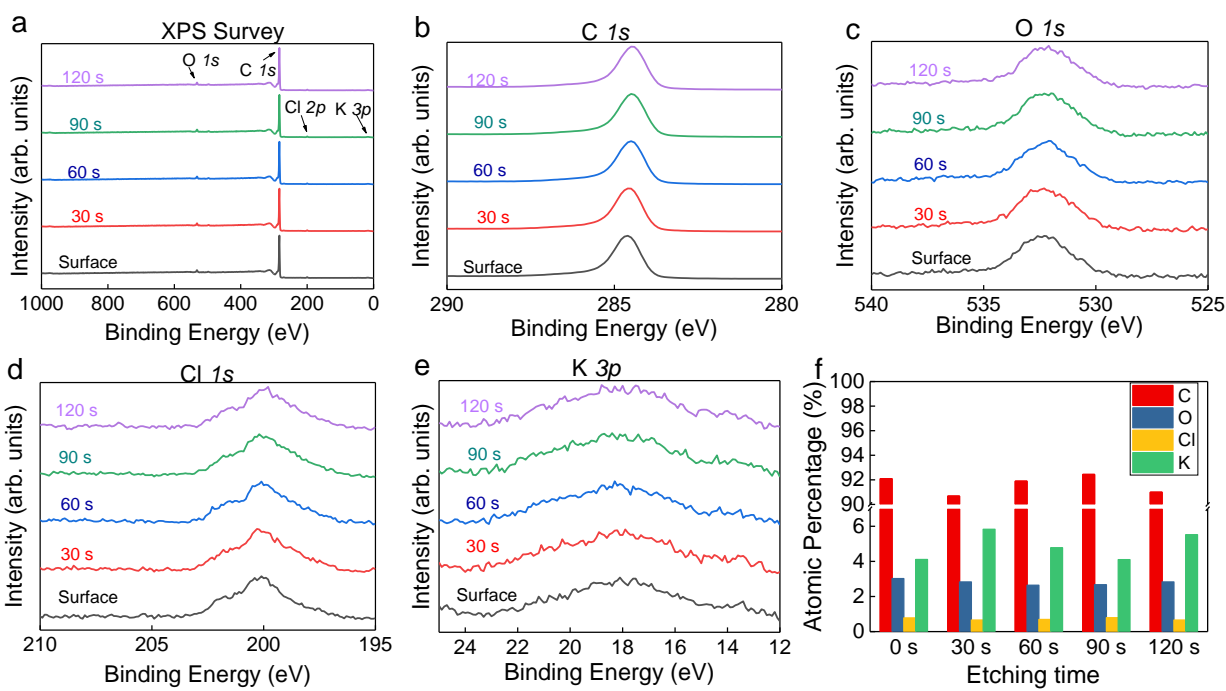

**Supplementary fig. 8: XPS depth analysis.** XPS data was collected at different depths of the intercalated graphite samples. XPS survey (a) and depth profiling of several elements found in ICG (b-e). (f) The atomic percentage of various elements present in ICG at different levels of depth.

**Section 9. Intercalation with water:**

We have carried out a few other tests to see the effect of intercalant DI water on the expansion of the interlayer spacing of graphite and to see if the epoxy interface is stable after applying a high voltage for a long time. For this purpose, we used water as the intercalant instead of KCl and applied a voltage of 10 V for three hours. We utilized this sample for ionic conductance measurements at different concentrations of NaCl. In contrary to the samples intercalated with KCl, the water-intercalated samples exhibited very little enhancement in the current (Supplementary fig. 9). This experiment strongly suggests the importance of the large concentration of ions for successful intercalation. The negligible increase in current after the water intercalation also confirms the stability of epoxy glue. It indicates that even after applying a high voltage, there is no significant leakage in the samples.

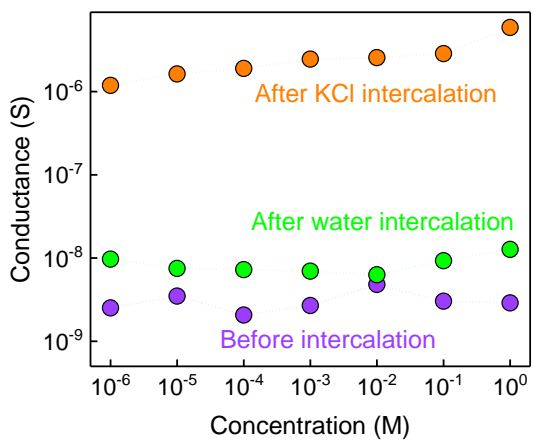

**Supplementary fig. 9: Graphite intercalation with water and KCl.** Comparison of conductance of graphite before and after the intercalation with water only and KCl only at several concentrations of NaCl. There is a clear enhancement in the ionic conductance through KCl intercalated samples, suggesting the importance of KCl in expanding the interlayer space.

**Supplementary references:**